# *BinPo*: An open-source code to compute the band structure of two-dimensional electron systems

Martínez E. A.[a,*], Beltrán Fínez J. I.[a], Bruno F. Y.[a]

[a] GFMC, Departamento de Física de Materiales, Universidad Complutense de Madrid, 28040 Madrid, Spain

## Abstract

We introduce *BinPo*, an open-source Python code to compute electronic properties of two-dimensional electron systems. Its usage is focused on the $ABO_3$ perovskite structure based systems, such as $SrTiO_3$ and $KTaO_3$, because of their increasing impact in materials community and possible applications in spintronic devices. *BinPo* has a Schrödinger-Poisson solver to obtain the self-consistent potential energy in a slab system. The tight binding slab Hamiltonian of the system is created from the transfer integrals in the maximally localized Wannier functions basis, thus reaching a higher accuracy than conventional tight binding methods. The band structure, energy slices, and other properties, along with different projections and orientations can be computed. High resolution and publishable figures of the simulations can be generated. In *BinPo*, priority has been given to ease-of-use, efficiency, readability and modularity, therefore becoming suitable to produce reliable electronic structures simulations at low computational cost. Along with the code itself, we provide files from first-principles calculations, instructions of use and detailed examples of its wide range of capabilities. We detail the approaches used in the code, so that it can be further exploited and adapted to other problems, such as adding new materials and functionalities which can strength the initial code scopes.

\* Corresponding author.

E-mail address: emanuelm@ucm.es (E. A. Martínez).

## 1. Introduction

In the year 2004 it was discovered that a high mobility two-dimensional electron system (2DES) is formed at the interface between the two insulating oxides LaAlO$_3$ (LAO) and SrTiO$_3$ (STO) [1]. This fascinating finding gave rise to an intense research effort in the oxide electronics community to better understand the underlying mechanisms, variations in physical properties and prospective applications of the 2DES. Later on, many different procedures for creating a 2DES were revealed, such as photon irradiation under high-vacuum of STO or electrolyte-gating the oxide surface, among others [2]. The fact that STO is easily doped by different methods makes difficult to elaborate a general theory explaining the origin of the 2DES in the vast amount of cases found [3]. Whereas in some cases purely electrostatic explanations are invoked, the presence of defects such as oxygen vacancies, appears of fundamental importance in others [4–7]. That is why it is remarkable that the electronic structure measured in different 2DES systems based on STO is almost indistinguishable [8–11]. Moreover, many other similarities are found in transport properties of different 2DES all based on STO crystals [12,13] and theoretical band structure calculations with different approaches [14–22].

Here we present a computational tool that allows for the calculation of the 2DES band structure in a variety of oxide-based systems that is independent of the microscopic origin of the charge carriers. The reader interested in the theoretical details of this dependency is referred to [23]. The computational tool relies on the band bending potential that can be obtained by solving the Schrödinger-Poisson scheme [24]. This produces self-consistent (SC) solutions comprising a distribution of the charge and potential energy along the confinement direction, enabling the computation of many electronic properties. *BinPo,* which stands for tight **bin**ding-**Po**isson calculations, is an open-source Python code to compute the band structure of oxide-based 2DES, where importance is given to ease-of-use and capability of reproducing direct measurements. Because of its simplified structure it is possible to use the code to calculate the electronic properties of a 2DES based on other oxides than STO, such as for example KTaO$_3$ (KTO). Calculations based on the same ideas were already presented in several publications in order to explain direct experimental measurements of the electronic structure of oxide based 2DES [8,10,25–27].

A few codes exist that perform similar tasks in conventional semiconductors: NextNano [28] which is widely used to simulate nanodevices with realistic geometries, 1D Poisson [29] employed for calculating energy band diagrams, and more recently, Aestimo1D [30], a program that allows simulating heterostructures in one dimension. All of them are Poisson-Schrödinger solvers allowing for the calculation of the SC potential energy and many other related properties in such semiconductor systems. To the best of our knowledge none are yet available for studying oxide compounds with strong electric field dependence of the relative permittivity as in the case of quantum paraelectric STO and KTO. Our contribution here is to introduce a ready-to-use code that allows for the calculation of the band structure in quantum confined 2DES stabilized in oxides. The already existing codes assume linear media for the Poisson equation, which is proper for semiconductors, in contrast *BinPo* directly solves the non-linear Poisson equation with a plane-by-plane discretization. Moreover, *BinPo* can take any user-defined relative permittivity model as an input to solve the Schrödinger-Poisson scheme. Some features included in *BinPo* beyond the computation of the band structure along the high symmetry paths are: visualization of SC solutions for the electronic density and confinement potential, energy slices calculations and the option to obtain orbitally and plane projected band structures, among others. Notably, since the starting point is a relativistic density functional theory (DFT) calculation, *BinPo* allows for the study of the Rashba effect at oxide 2DES [25,27,31]. All the functionalities and the implementation in the *BinPo* code are explained and exemplified below.

The manuscript is organized as follows: in Section 2 we describe the background and mathematical methods used, in Section 3 we describe the program and the main parameters, in Section 4 we show step by step examples of SC potential, band structure calculations and the other facilities. We compare some simulations to experimental results. Finally, in Section 5 we summarize the work and present the conclusions.

## 2. Background and methods

The starting point to perform *BinPo* calculations is the Hamiltonian of a bulk system in the basis of maximally localized Wannier functions (MLWF) [32,33]. Although several semi-empirical approaches exist to generate tight binding (TB) models, such as *k.p* perturbation theory [34] and Slater-Koster method [35], it has been shown that the MLWF act as an exact TB basis capturing all electronic features from first-principles calculations [33,36]. TB models from MLWF basis, or simply MLWF TB models, are widely used to obtain realistic results and accurately compute physical quantities [36]. In this section we will show the methodology used to solve the quantum-electrostatic problem in the system starting from the bulk Hamiltonian, and subsequently generating a slab Hamiltonian to add a potential energy term. Despite the generality used to describe the methodology, at this stage, *BinPo* is programmed to work with ABO$_3$ cubic perovskites for the main crystallographic directions using the $t_{2g}$ manifold, that is $d_{xy}$, $d_{yz}$ and $d_{xz}$ orbitals [37].

### 2.1. The tight binding Hamiltonian of bulk systems

The use of MLWF basis enables to construct a set of Bloch-like states by performing a Fourier transform. Let **R** label the 3D real lattice vectors and $\alpha$ be the orbital index corresponding to a specific band. The transformation into k-space can be written as:

$$|\mathcal{W}_{\mathbf{k}\alpha}\rangle = \frac{1}{\sqrt{N}} \sum_{\mathbf{R}} e^{i\mathbf{k}\cdot\mathbf{R}} |\mathbf{R}\alpha\rangle \qquad (1)$$

where the states $|\mathcal{W}_{\mathbf{k}\alpha}\rangle$ are the Bloch elements, $|\mathbf{R}\alpha\rangle$ are the elements of the MLWF basis and $N$ is the number of mesh points in the first Brillouin zone. The orthonormalization of the MLWF basis, $\langle \mathbf{R}'\alpha'|\mathbf{R}\alpha\rangle = \delta_{\mathbf{R}\mathbf{R}'}\delta_{\alpha\alpha'}$, next to the convention used in Eq. (1) implies that in the reciprocal space the orthogonality reads $\langle \mathcal{W}_{\mathbf{k}'\alpha'}|\mathcal{W}_{\mathbf{k}\alpha}\rangle = \delta_{\mathbf{k}\mathbf{k}'}\delta_{\alpha\alpha'}$. Typically, the states $|\mathcal{W}_{\mathbf{k}\alpha}\rangle$ are not eigenstates of the Hamiltonian, therefore the Hamiltonian matrix will be non-diagonal. The transfer integrals are found to be [33]:

$$H^{(\mathbf{k})}_{\alpha\alpha'} = \langle \mathcal{W}_{\mathbf{k}\alpha'} | \hat{\mathcal{H}}_{KS} | \mathcal{W}_{\mathbf{k}\alpha} \rangle = \sum_{\mathbf{R}} e^{i\mathbf{k}\cdot\mathbf{R}} H^{(\mathbf{R})}_{\alpha\alpha'} \qquad (2)$$

where $\hat{\mathcal{H}}_{KS}$ is the Kohn-Sham (KS) Hamiltonian of first-principles calculations and $H^{(\mathbf{R})}_{\alpha\alpha'} = \langle 0\alpha' | \hat{\mathcal{H}}_{KS} | \mathbf{R}\alpha \rangle$ are the elements of the KS Hamiltonian projected onto the MLWF basis. Note that the Schrödinger equation can be now solved fast and straightforwardly by diagonalization to get the band structure of the bulk system.

The procedure to generate the MLWF basis on top of first-principles calculations is the so-called Wannierization. To carry out this procedure we use the open-source Wannier90 code [38–40], which is a standard tool in the condensed matter community since it is interfaced within most of the commonly used first-principles programs, such as VASP, WIEN2k, Quantum Espresso, Abinit and SIESTA among others. While any of the programs mentioned produces the needed output files to generate the MLWF basis, the examples in this paper were obtained from Quantum Espresso [41]. The details of these calculations, as well as the details of the Wannierization, can be found in Appendix A.

The workflow to obtain the $H^{(\mathbf{R})}_{\alpha\alpha'}$ elements is the following: perform first-principles calculations, get the MLWF basis by Wannierization using the $t_{2g}$ manifold and then projecting the KS Hamiltonian onto this basis. In Wannier90 software the $H^{(\mathbf{R})}_{\alpha\alpha'}$ elements are saved to the *seedname_hr.dat* file and henceforth we refer to it as W90 file, which is provided for the cubic STO and KTO in this work. This file will be taken by *BinPo* to generate the slab Hamiltonian in k-space described in the next subsection.

**2.2. The tight binding slab Hamiltonian**

In order to extend the previous discussion of the bulk Hamiltonian to a specific slab model, we stablish a real space direction, normal to the desired slab face, with a unit vector $\hat{\mathbf{r}}_\perp$. Then, we separate the 3D real vectors into a longitudinal ($\mathbf{R}_\perp$) and transverse ($\mathbf{R}_\parallel$) contributions. So that $\mathbf{R} = \mathbf{R}_\parallel + \mathbf{R}_\perp$, with $\mathbf{R}_\perp = r_\perp \hat{\mathbf{r}}_\perp$, where $r_\perp$ is an integer index labelling the system plane and $\mathbf{R}_\parallel$ includes the lattice vectors perpendicular to $\hat{\mathbf{r}}_\perp$. In this way, in analogy to Eq. (2), the Fourier transform for a slab can be denoted as:

$$H^{(\mathbf{k}_\parallel)}_{\alpha\alpha'r_\perp r'_\perp} = \langle \mathcal{W}_{\mathbf{k}_\parallel \alpha' r'_\perp} | \hat{\mathcal{H}}_{KS} | \mathcal{W}_{\mathbf{k}_\parallel \alpha r_\perp} \rangle = \sum_{\mathbf{R}_\parallel} e^{i\mathbf{k}_\parallel \cdot \mathbf{R}_\parallel} \langle 0\alpha' r'_\perp | \hat{\mathcal{H}}_{KS} | \mathbf{R}_\parallel \alpha r_\perp \rangle \qquad (3)$$

where now $|\mathbf{R}_\parallel \alpha r_\perp\rangle$ states are the elements of the MLWF basis discretized by planes, which in turn have the associated $|\mathcal{W}_{\mathbf{k}_\parallel \alpha r_\perp}\rangle$ states in the k-space. The above formula is nothing other than the 2D Fourier transform at each plane applied on the projected KS Hamiltonian. The orthogonality relations in the real and reciprocal space are:

$$\langle \mathbf{R}'_\| \alpha' r'_\perp | \mathbf{R}_\| \alpha r_\perp \rangle = \delta_{\mathbf{R}_\| \mathbf{R}'_\|} \delta_{\alpha \alpha'} \delta_{r_\perp r'_\perp}$$
$$\langle \mathcal{W}_{\mathbf{k}'_\| \alpha' r'_\perp} | \mathcal{W}_{\mathbf{k}_\| \alpha r_\perp} \rangle = \delta_{\mathbf{k}_\| \mathbf{k}'_\|} \delta_{\alpha \alpha'} \delta_{r_\perp r'_\perp} \quad (4)$$

These orthogonality relations are exact, if and only if, the discretization keeps the volume of the bulk unit cell unchanged. Otherwise, these relations are still good approximations (See the discussion in Appendix E).

So far, the Hamiltonian matrix will be a block diagonal one with translation symmetry along the $\hat{\mathbf{r}}_\perp$ axis and it would range all over the space for $r_\perp \to \pm \infty$. Nevertheless, we are interested in the region of the space containing the crystal. In practice, the index $r_\perp$ must be truncated to model such a region. Let $L$ be the number of planes in the slab, we state that the transfer integrals are strictly zero if any of $r_\perp, r'_\perp$ is outside of the $[0, L-1]$ range. The Hamiltonian matrix generated by this constraint, which we call the slab Hamiltonian matrix, is:

$$H^{(\mathbf{k}_\|)} = \begin{pmatrix} H^{(\mathbf{k}_\|)}_{\alpha\alpha'00} & H^{(\mathbf{k}_\|)}_{\alpha\alpha'01} & H^{(\mathbf{k}_\|)}_{\alpha\alpha'02} & & H^{(\mathbf{k}_\|)}_{\alpha\alpha'0,L-1} \\ H^{(\mathbf{k}_\|)}_{\alpha\alpha'10} & H^{(\mathbf{k}_\|)}_{\alpha\alpha'11} & H^{(\mathbf{k}_\|)}_{\alpha\alpha'12} & \cdots & H^{(\mathbf{k}_\|)}_{\alpha\alpha'1,L-1} \\ H^{(\mathbf{k}_\|)}_{\alpha\alpha'20} & H^{(\mathbf{k}_\|)}_{\alpha\alpha'21} & H^{(\mathbf{k}_\|)}_{\alpha\alpha'22} & & H^{(\mathbf{k}_\|)}_{\alpha\alpha'2,L-1} \\ & \vdots & & \ddots & \vdots \\ H^{(\mathbf{k}_\|)}_{\alpha\alpha'L-1,0} & H^{(\mathbf{k}_\|)}_{\alpha\alpha'L-1,1} & H^{(\mathbf{k}_\|)}_{\alpha\alpha'L-1,2} & \cdots & H^{(\mathbf{k}_\|)}_{\alpha\alpha'L-1,L-1} \end{pmatrix} \quad (5)$$

where the main diagonal blocks $H^{(\mathbf{k}_\|)}_{\alpha\alpha' r_\perp r_\perp}$ correspond to the intraplane interactions, while the off-diagonal blocks $H^{(\mathbf{k}_\|)}_{\alpha\alpha' r_\perp r'_\perp}$ correspond to interplane ones. It should be noted that, according to Eq. (3), each block has dimension $N_\alpha \times N_\alpha$, with $N_\alpha$ the number of elements in the MLWF basis. Therefore, the slab Hamiltonian matrix will have a dimension equal to $N_\alpha L \times N_\alpha L$ at each $\mathbf{k}_\|$ point. In the specific problem treated here, $\alpha$ labels the $t_{2g}$ manifold, *i. e.*:

$$\alpha = \{d^\uparrow_{yz}, d^\downarrow_{yz}, d^\uparrow_{zx}, d^\downarrow_{zx}, d^\uparrow_{xy}, d^\downarrow_{xy}\} \quad (6)$$

We already made the implicit assumption here that the MLWF can be thought of as *d*-orbitals in the crystal. This is justified because these functions are centered on the atomic nuclei and clearly exhibit an atomic orbital character [42]. Consequently, we will talk indistinguishably about the MLWF of the $t_{2g}$ manifold and the *d*-orbitals actually constituting it.

Since the goal is to obtain the electrostatic potential energy for the system, hereafter the SC potential, we take advantage of this expanded Hamiltonian matrix to add an on-site potential energy term. This potential energy is supposed to be homogeneous in the plane and should vary smoothly along $\hat{\mathbf{r}}_\perp$ axis. Later, this potential will lead to the possibility of obtaining a confined charge profile at the surface by solving the quantum-electrostatic problem in the slab. The slab Hamiltonian matrix with the potential term added reads:

$$\begin{pmatrix} H^{(\mathbf{k}_{||})}_{\alpha\alpha'00} + \underline{V}(0) & H^{(\mathbf{k}_{||})}_{\alpha\alpha'01} & \cdots & H^{(\mathbf{k}_{||})}_{\alpha\alpha'0,L-1} \\ H^{(\mathbf{k}_{||})}_{\alpha\alpha'10} & H^{(\mathbf{k}_{||})}_{\alpha\alpha'11} + \underline{V}(1) & & H^{(\mathbf{k}_{||})}_{\alpha\alpha'1,L-1} \\ \vdots & \vdots & \ddots & \vdots \\ H^{(\mathbf{k}_{||})}_{\alpha\alpha'L-1,0} & H^{(\mathbf{k}_{||})}_{\alpha\alpha'L-1,1} & \cdots & H^{(\mathbf{k}_{||})}_{\alpha\alpha'L-1,L-1} + \underline{V}(L-1) \end{pmatrix} \quad (7)$$

where the quantity $\underline{V}(r_\perp)$ represents the $N_\alpha \times N_\alpha$ potential energy matrix which affects each plane. We assume that this potential energy is applied on all the on-site elements in the same way, despite their orbital character and without promoting any electronic inter-orbital transition. In consequence, $\underline{V}(r_\perp) = V(r_\perp) \cdot \mathbb{1}_{N_\alpha \times N_\alpha}$, being $V(r_\perp)$ the potential energy per plane and $\mathbb{1}_{N_\alpha \times N_\alpha}$ the $N_\alpha \times N_\alpha$ identity matrix. A sketch of the system modeled by the matrix of Eq. (7) is depicted in Fig. 1.

Finally, we could write the TB slab Hamiltonian operator for this system in second quantization as:

$$\widehat{\mathcal{H}}_{TB} = \sum_{\mathbf{k}_{||}\alpha\alpha'r_\perp r_\perp'} \left( H^{(\mathbf{k}_{||})}_{\alpha\alpha'r_\perp r_\perp'} + \delta_{\alpha\alpha'}\delta_{r_\perp r_\perp'} V(r_\perp) \right) c^\dagger_{\mathbf{k}_{||}\alpha r_\perp} c_{\mathbf{k}_{||}\alpha' r_\perp'} \quad (8)$$

where $c^\dagger_{\mathbf{k}_{||}\alpha r_\perp}$ and $c_{\mathbf{k}_{||}\alpha r_\perp}$ are the creation and annihilation operators of one electron in the state $|\mathcal{W}_{\mathbf{k}_{||}\alpha r_\perp}\rangle$.

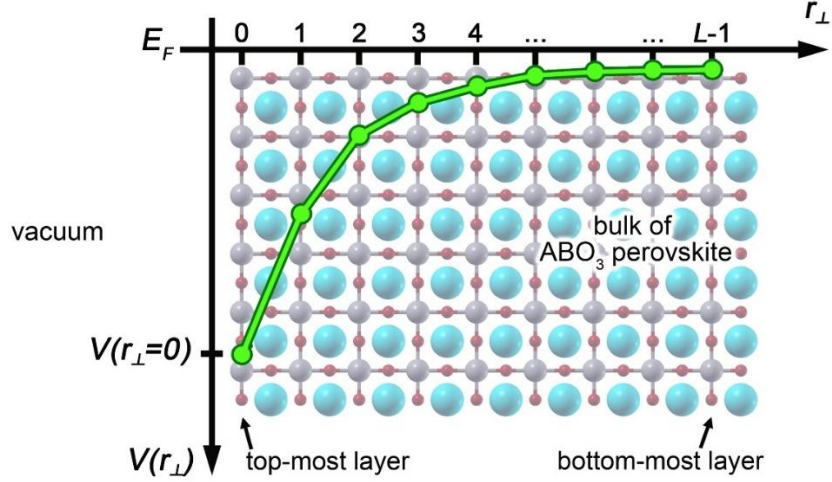

**Fig 1.** Schematic picture of the system modelled by the slab Hamiltonian matrix with an additional on-site potential energy term.

### 2.3. The Schrödinger-Poisson scheme

To find out what is the SC potential energy of the system, the quantum-electrostatic problem using the well-known Schrödinger-Poisson scheme [24] must be solved. Our approach follows earlier work described in Refs. [43–45]. The main assumptions we make are the following: (a) the crystal presents no defects and no structural relaxations, (b) electrons are uniformly distributed within a well-defined volume between adjacent planes, (c) after applying boundary conditions for the potential, the quantum-electrostatic problem is uniquely determined by the magnitude of the potential at the top-most layer, $V(0)$, and (d) the effect of polar distortions, without the need of establishing microscopic mechanisms, are taken into account within the permittivity model. This last assumption is based on some of the arguments given in Ref. [22]. Note that in the case of STO, the effects of tetragonal distortions are neglected [46].

The Schrödinger-Poisson scheme solves self-consistently the potential energy $V(r_\perp)$ in the Schrödinger, charge density and Poisson equations. Then, a given convergence criterion is satisfied by obtaining, after a number of iterations, the output ($V_{out}$) sufficiently close to the input ($V_{in}$). Now, each of these 4 procedures are described.

### 2.3.1. The tight binding step

The time-independent Schrödinger equation for the slab TB Hamiltonian operator of Eq. (8) reads:

$$\widehat{\mathcal{H}}_{TB} |\phi_{\mathbf{k}_\parallel n}\rangle = \epsilon_{\mathbf{k}_\parallel n} |\phi_{\mathbf{k}_\parallel n}\rangle \quad (9)$$

As in the bulk case, this equation is solved straightforwardly by diagonalizing the slab Hamiltonian matrix. Then, $N_\alpha L$ eigenenergies $\epsilon_{\mathbf{k}_\parallel n}$ are found at every $\mathbf{k}_\parallel$-point in reciprocal space. For each eigenenergy (enumerated by $n$ index) there is an associate eigenstate, which is of dimension $N_\alpha L$. The obtained eigenstates can be expressed in terms of a set of complex coefficients $u_{\alpha r_\perp}^{\mathbf{k}_\parallel n}$ as:

$$|\phi_{\mathbf{k}_\parallel n}\rangle = \sum_{\alpha, r_\perp} u_{\alpha r_\perp}^{\mathbf{k}_\parallel n} |\mathcal{W}_{\mathbf{k}_\parallel r_\perp}\rangle \quad (10)$$

By applying the orthogonality relations of Eq. (4), coefficients $u_{\alpha r_\perp}^{\mathbf{k}_\parallel n}$ must satisfy the following normalization condition:

$$\sum_{\alpha, r_\perp} \left|u_{\alpha r_\perp}^{\mathbf{k}_\parallel n}\right|^2 = 1 \quad (11)$$

### 2.3.2. Charge density calculation

By means of the orthogonality relations given in Eq. (4), the charge density can be computed in term of the coefficients $u_{\alpha r_\perp}^{\mathbf{k}_\parallel n}$ as follows:

$$\rho(r_\perp) = -\frac{|e|}{\Omega_\perp N_\parallel} \sum_{\mathbf{k}_\parallel n} f(\epsilon_{\mathbf{k}_\parallel n}, T) \sum_\alpha \left|u_{\alpha r_\perp}^{\mathbf{k}_\parallel n}\right|^2 \quad (12)$$

where $|e|$ is the elementary charge, $N_\parallel$ is the total number of k-points used to discretize the first 2D Brillouin zone (BZ1), $f(\epsilon_{\mathbf{k}_\parallel n}, T)$ is the Fermi-Dirac function evaluated on the eigenenergy $\epsilon_{\mathbf{k}_\parallel n}$ at a temperature $T$, and $\Omega_\perp$ is the interplanar volume over which the electrons are uniformly distributed and it is defined as:

$$\Omega_\perp = \|\bar{a}_1 \times \bar{a}_2\| \Delta r_\perp \quad (13)$$

where $\bar{a}_1$, $\bar{a}_2$ are the lattice vectors contained in $\mathbf{R}_\parallel$ and $\Delta r_\perp$ is the interplanar distance along the $\hat{\mathbf{r}}_\perp$ axis between two contiguous planes. Further details on the charge density formula and its validity are discussed in Appendix E.

### 2.3.3. The Poisson step

The non-linear Poisson equation as derived from the Gauss law is:

$$\frac{d^2 V(r_\perp)}{dr_\perp^2} = \frac{|e|}{\varepsilon_0 \varepsilon_r[E(r_\perp)]} \rho(r_\perp) \tag{14}$$

where $\varepsilon_0$ is the vacuum permittivity and $\varepsilon_r$ is the relative permittivity, which is a functional of the electric field, $E$. This introduces the non-linearity in the Poisson equation because $E$, in turn, depends on the potential energy as $E(r_\perp) = \frac{1}{|e|}\frac{dV}{dr_\perp}(r_\perp)$. Moreover, due to the quantum paraelectric nature of the involved ABO$_3$ perovskites [47–50], the relative permittivity strongly affects the potential energy profile along the slab. Some comments of the relative permittivity models used here can be found in Appendix B.

Eq. (14) is a second order differential equation, so that we must explicit two boundary conditions. In our approach, a Dirichlet boundary condition is used at the top-most layer, whereas either a Dirichlet or a Neumann one can be used at the bottom-most layer.

After solving the non-linear Poisson equation, we obtain an output potential energy ($V^{out}$) which can be compared with the input one to ensure electronic convergence, see Subsection 2.3.4. Further information on our approach to obtain and solve Eq. (14), as well as the implementation of boundary conditions, is provided in Appendix E.

### 2.3.4. Convergence criterion

The difference between the input and output potential energies obtained from the Poisson equation is calculated according to:

$$\chi^2 = \frac{1}{L} \sum_{r_\perp} \left( \frac{V^{out}(r_\perp) - V^{in}(r_\perp)}{V(r_\perp = 0)} \right)^2 \tag{15}$$

If this value is less than a defined convergence threshold at the $j$-th iteration, the SC solution is considered found. On the other hand, if this criterion is not satisfied, the under-relaxation mixing algorithm [43,51] generates a new input potential for the $(j+1)$-th iteration as:

$$V_{in}^{j+1}(r_\perp) = V_{in}^j(r_\perp) + f_{mix}\left(V_{out}^j(r_\perp) - V_{in}^j(r_\perp)\right) \tag{16}$$

with $f_{mix}$ being the mixing factor. Generally, self-consistency is achieved for $f_{mix}$ between 0.05-0.3 [43]. A flowchart of the whole Schrödinger-Poisson scheme is presented in Fig. 2.

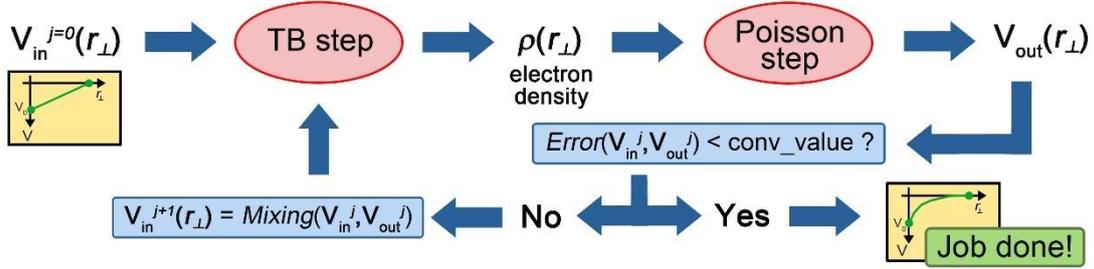

**Fig 2.** Flowchart of the Schrödinger-Poisson scheme.

Now, we will follow the scheme shown in Fig. 2 to summarize the algorithm. In the very first step (iteration $j = 0$) the initial potential energy and the boundary conditions are set. The initial potential is arbitrarily set to either a triangular or exponential functional form. Then, a self-consistency loop is executed along the following steps at the $j$-th iteration:

(i) TB step: solve the Schrödinger equation for a potential energy $V_{in}^j$. This potential is properly added to the slab Hamiltonian matrix as indicated in Eq. (7), then the matrix is diagonalized to find the corresponding eigenvalues and eigenvectors.
(ii) Compute the electron density from the resulting eigenvalues and eigenvectors using Eq. (12).
(iii) Poisson step: solve the 1D non-linear Poisson equation (Eq. (14)) to get a new potential energy $V_{out}^j$.
(iv) Convergence criterion: compare if $V_{in}^j$ and $V_{out}^j$ are equal up to a tolerance value according to the error value obtained in Eq. (15), else generate the input potential for the $(j + 1)$-th iteration using Eq. (16) and return to point (i).

## 2.4. Post-processing steps

Once the quantum-electrostatic problem is solved, we have access to the electronic density and potential energy as a function of plane index satisfying some bulk value, inferred from DFT, and certain value at the surface. The $V(r_\perp)$ potential energy in the slab Hamiltonian operator of Eq. (8) is then replaced by the found SC potential. The eigenenergies and eigenstates coefficients obtained by solving this Schrödinger equation (Eq. (9)) provide important features of the system. This typically consists of specific quantities projected onto any of the system phase space dimensions (reciprocal k-points, orbitals of the manifold, plane sites, etc.). For example, the band structure with different projections, the Fermi surface, or the electron density with orbital contributions among other quantities can be computed. Appendix C summarize the quantities that were included as post-processing components of *BinPo*, with the respective derivations. On the other hand, in Section 4 we will show how to execute each of the post-processing components.

**3. Software description**

     *BinPo* was programmed in Python (version 3.x), which has become one of the most popular programming languages, especially for data science [52]. The libraries required to run the code are the standard ones NumPy [53], SciPy [54] and Matplotlib [55]. *BinPo* also uses the Atomic Simulation Environment (ASE) [56], which is a powerful Python library to perform atomistic simulations.

     Firstly, *BinPo* requires to read a W90 file based on a relativistic first-principles calculation, whose main advantage is that all spin-orbit interactions are extracted from the original W90 input file. Therefore, spin splitting and the associated phenomena like avoided crossing [57–59] or Rashba effect [25,60,61] can be obtained as a direct consequence of the model. Despite that we already provide W90 input files for STO and KTO with *BinPo*, the user could append other W90 files, with different first-principles treatments and/or different $ABO_3$ materials. This will take full advantage of the MLWF TB model of the slab by capturing specific features from DFT calculations. Importantly, the DFT calculation and the Wannierization must be done in a cubic cell, which significantly reduces the computational cost. Then, if necessary to calculate properties on 2DES confined along directions $\mathbf{r}_\perp$ different than [001], an internal algorithm in *BinPo* allows for rotating the reference system. This algorithm will be described in Appendix E.

     In the *BinPo* folder the user will find several Python files, which we name components. First of all, the user must execute the *BP-preproc.py* component for a pre-processing step. The working structure of this step can be seen in Fig. 3(a). This pre-processing consists of performing a separation of the matrix elements of the MLWF TB Hamiltonian into planes along the $\hat{\mathbf{r}}_\perp$ axis. The *BP-preproc.py* component calls some standard Python modules and the *BPdatabase.py* module, loads the configuration and the W90 files and two parameters: the material name and the 2DES confinement direction (as *hkl* indices). The *BPdatabase.py* module contains the information arising from DFT calculations, like the lattice parameters and the lowest unoccupied levels. In the current version of *BinPo* allowed materials are STO y KTO, while allowed crystallographic directions are [100], [110] and [111]. After a successful pre-processing, the slab Hamiltonian for a specific material along certain direction will appear in the *BinPo* folder under the name *Hr+material+hkl*.

     The next step is to perform the SC potential calculation, as indicated by the workflow in Fig. 3(b). By running the *BP-scp.py* component, the following steps will carry on: necessary modules (numpy, matplotlib, BPmodule.py, ASE, etc.) are loaded, the parameters are set from the configuration files and/or the command-line, the files in *Hr+material+hkl* folder are loaded and calculation is done. Every file (including a *.log* file) are saved to a new folder with an exclusive name, the identifier, defined by the user

for every calculation. Each post-processing step after this point will call the files inside the output folder of the SC potential calculation without modifying its content, except for the *.log* file. For instance, the band structure with *BP_bands.py* or the Fermi surface with *BP_energy-slices.py* can be computed and the outputs will be saved to the respective identifier folder. The details about the use of *BinPo* can be found in *README.md* file within the *BinPo* folder. The capabilities of *BinPo* through the different components are summarized in Table 1. Each component has an associated configuration file (except for *BP-preproc.py* and *BP-fast_plot.py*) for reading default values for the parameters in case the user does not provide them by command-line.

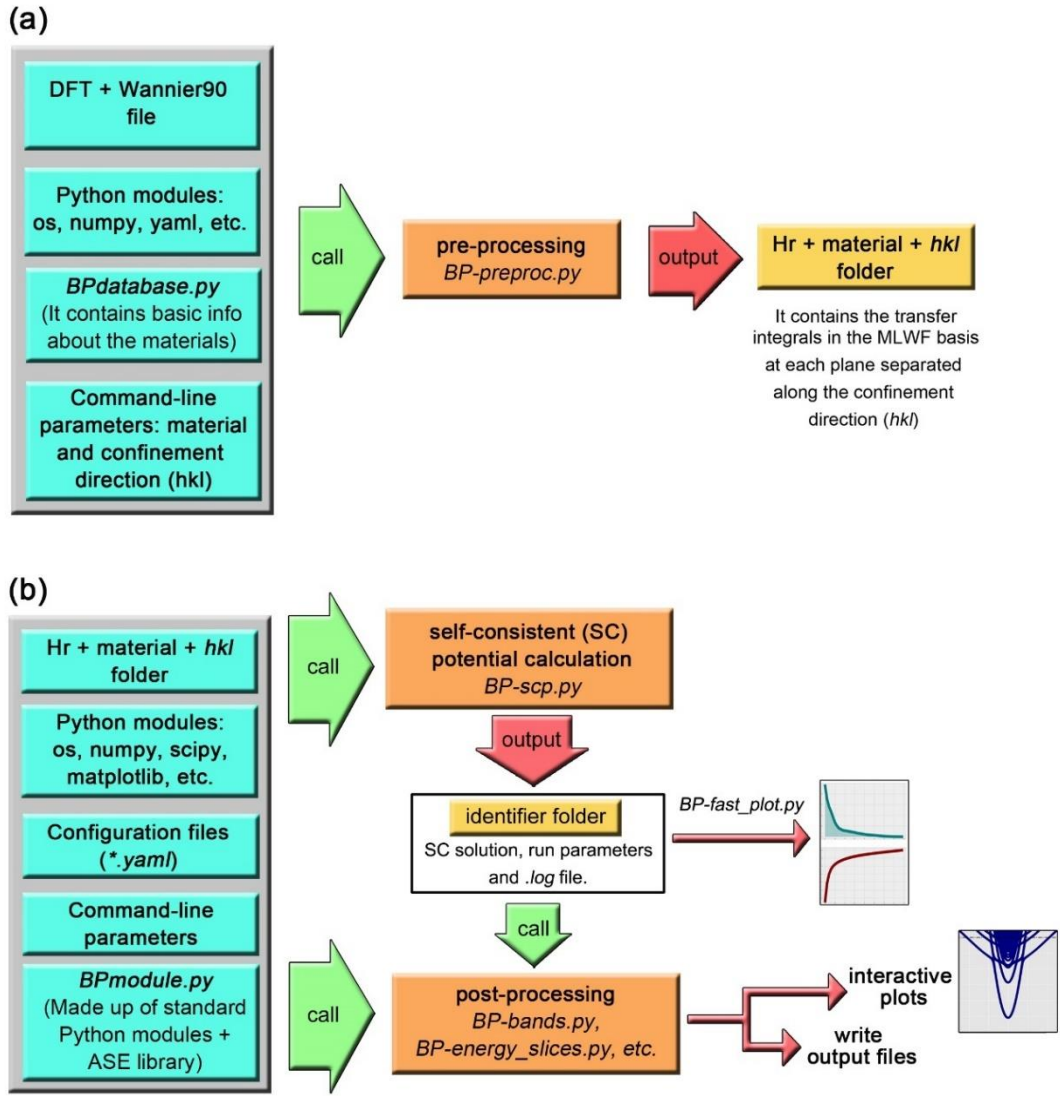

**Fig 3.** Working structure of *BinPo*. (a) Pre-processing step. (b) Self-consistent (SC) potential energy calculation and post-processing steps.

In Fig. 3(b) can be seen that *BinPo* calls a calculation module named *BPmodule.py*. This is the main module holding the classes and methods. In Section 2.2 the concepts of the slab Hamiltonian and the potential energy along the slab were introduced. These quantities find their counterparts with the Quasi2DHamiltonian *class* and the PotentialEnergy *class* inside this module. These classes contain the fundamental methods to execute the Schrödinger-Poisson scheme. There is a third class called CrystalFeatures, which addresses the crystallography issues of the system. In the next section we will show a complete example of how to use *BinPo* from command-line to compute several quantities of a STO(100) slab.

Table 1. *BinPo* capabilities: components and their functions.

| Component | Description |
|---|---|
| **Pre-processing and SC potential energy calculation** | |
| *BP-preproc.py* | Pre-processing component. It generates the *Hr+material+hkl* folder to be loaded in the SC potential calculation and post-processing routines. It runs by command-line without an associated configuration file. |
| *BP-scp.py* | Main component. It performs the SC potential energy calculation. Config. file: *~/config_files/ scp.yaml*. |
| **Post-processing** | |
| *BP-fast_plotter.py* | Quick plotter tool for the output of *BP-scp.py*. It runs by command-line without an associated configuration file. |
| *BP-bands.py* | Band structure component. It performs a band structure calculation, which can include orbital character and/or bands projections onto planes. Config. file: *~/config_files/bands.yaml*. |
| *BP-energy_slices.py* | Energy slices component. It performs the calculation of any energy slice (*e. g.* Fermi surface). Config file: *~/config_files/energy_slices.yaml*. |
| *BP-energy_plot.py* | Plotter component for *BP-energy_slices.py* output. Config. file: *~/config_files/energy_plot.yaml*. |
| *BP-orb_density.py* | Component to decompose the electron density according to the orbital character. Config. file: *~/config_files/orb_density.yaml*. |
| *BP-envelope_wfs.py* | Component to compute the envelope wavefunctions at the $\Gamma$-point. Config. file: *~/config_files/envelope_wfs.yaml*. |

## 4. Examples

In this section, we will show an entire calculation example with *BinPo* for the archetypal system STO(100). However, the same steps could be followed to compute the physical quantities for different $ABO_3$ perovskites and/or other directions. Importantly, we have aimed to organize the input method in a modular manner, so that the user can enter parameter values by command-line, whereas the omitted values are read from the specific configuration files, see Table 1. Moreover, the configuration files are carefully structured to be easily customizable. To obtain from command-line a list of parameters that can be modified, the user should type:

$ *python BP-component.py -h*

being *BP-component.py* any of the components of *BinPo*. By doing so, a list of the basic updatable parameters will appear. The configuration file settings along with the command-line typing for all the examples presented in this section can be found in *~/BPexamples*.

## 4.1 Pre-processing step

Firstly, the user must perform the pre-processing step by typing in the command-line:

*$ python BP-preproc.py -mt STO -cfd 100*

Note that the material and the 2DES confinement direction, must be indicated by means of *mt* and c*fd* parameters, respectively. After finishing the pre-processing for a specific material/direction combination, the *Hr+material+hkl* folder will appear. It is important to note that this component should be run only once per material/direction combination. Subsequent SC potential calculations and post-processing steps will call the files from *Hr+material+hkl* folder.

## 4.2 Self-consistent potential energy calculation

Once the slab Hamiltonian is stored, calculations of the electrostatic properties, such as the SC potential energy, electron density, electric field, etc. can be obtained after specifying the slab size, boundary conditions and k-point grid sampling among other quantities. To achieve this, the user should run the SC potential energy calculation component as:

*$ python BP-scp.py -id run1 -mt STO -cfd 100 -tl 40 -nk 26 -bc1 -0.22*

In the above command-line we first indicate to *BinPo* that the identifier (*id*) is "run1". It will remain exclusive for this calculation and can be recalled in post-processing steps. The material (*mt*) is STO and the slab consists of 40 planes (*tl*) stacked along the (100) direction (*cfd*) and computed in a k-grid (*nk*) of 26 x 26 points. The boundary condition at the top-most layer (*bc1*) is -0.22 eV. The outputs of this calculation will be saved to *run1* folder. At the end of the calculation, the user will find a *.dat* file holding the SC solution, the *.log* and the *.yaml* configuration file with the parameters involved.

After the calculation is done, the user may try to vary the boundary condition value at the top-most layer, for example -0.36 eV, but this time the identifier (*id*) must be updated:

*$ python BP-scp.py -id run2 -mt STO -cfd 100 -tl 40 -nk 26 -bc1 -0.36*

Note that we modify uniquely the parameters that we need to update, while the other ones remain the same. Now the user will find another folder with the identifier *run2* inside the *BinPo* one. Both the potential energy and the electron density as a function of planes, plotted with *BP-fast_plot.py* tool, are shown in Fig. 4.

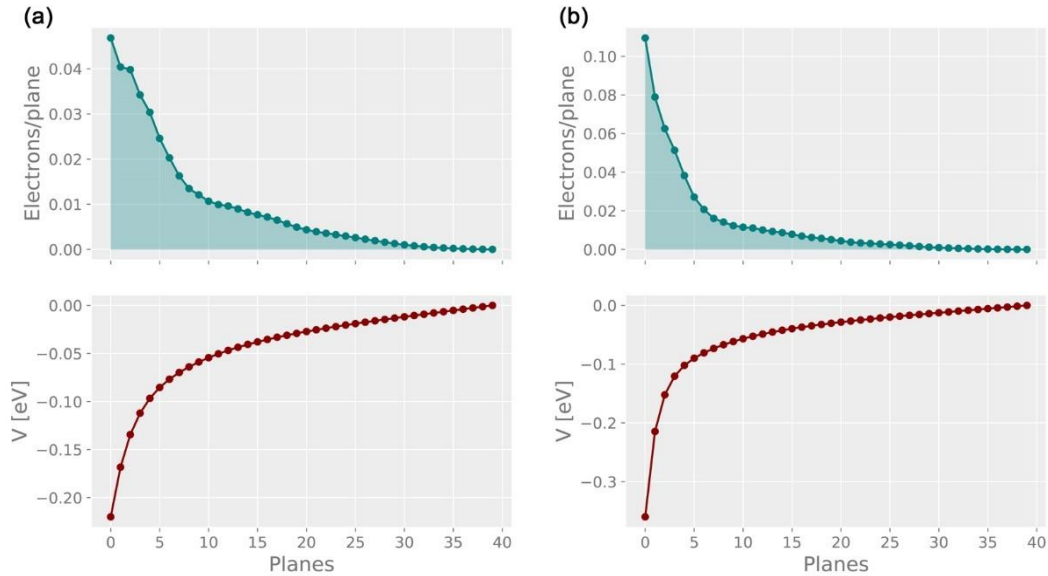

**Fig 4.** SC solutions of STO(100) (top: electron density, bottom: SC potential) with (a) bc1 = -0.22 eV and (b) bc1 = -0.36 eV boundary conditions at the top-most layer.

Once the user has done the SC potential energy calculation, each post-processing component can be executed independently. In the following sections we will show how to run such components and how the outputs look like.

### 4.3 Post-processing steps

Here, we will show how to apply post-processing steps on one of the SC solutions that we previously found.

#### 4.3.1 Band structure calculation

The *BP-bands.py* component can compute the band structure in several ways. The first one is what we call "total band structure", which corresponds to the electronic bands without including any projection. On the other hand, two different projections, onto orbitals or planes, can be computed as we will see below.

**Total band structure**

To compute the total band structure the user can type for example:

*$ python BP-bands.py -id run2 -ph XGX -kp 600 -tk 0 -nb 50*

Here, the path in the irreducible BZ1 between high symmetry points (*ph*) is specified to be X-Γ-X and it contains 600 points in k-space (*kp*). The task parameter (*tk*) equal to 0 indicates that band structure is calculated without projections and the number of bands (*nb*) to calculate will be 50. The user could also select a different path, for example:

*$ python BP-bands.py -id run2 -ph MGM -kp 600 -tk 0 -nb 50*

Both output plots are shown in Fig. 5(a) and (b).

*BinPo* also allows for simulating specific regions of interest in the band structure. For example, in order to closely examine the unconventional Rashba spin splitting, which emerges near the avoided crossings, the user could type:

$ *python BP-bands.py -id run2 -ph GX -kp 600 -tk 0 -nb 50 -xy 0.06 0.28 -0.15 -0.015*

In this case, a shorter path *ph* was selected (Γ-X), but the number of k-points was kept constant, thus a higher resolution zoomed image is obtained. Additionally, the *xy* parameter was introduced, which stablishes the x and y limit values of the energy-momentum plot window. The output of this calculation is shown in Fig. 5(c).

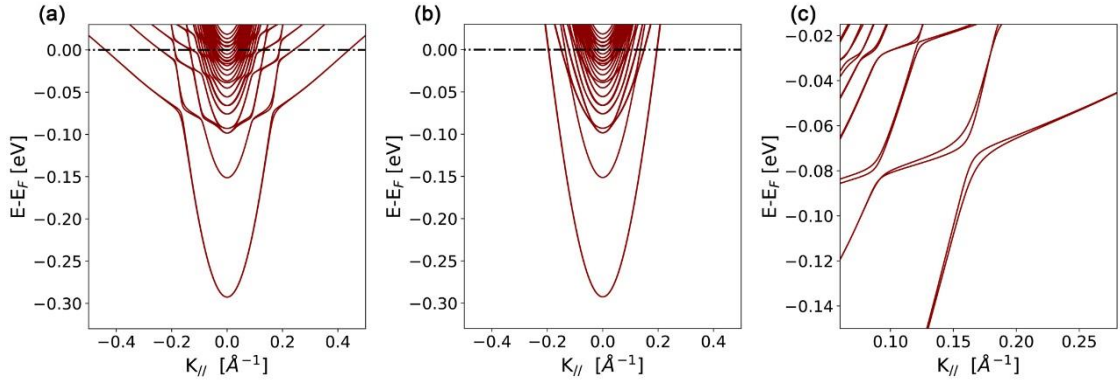

**Fig 5.** Output plots of total band structure calculation for STO(100) along (a) the X-Γ-X and (b) the M-Γ-M high symmetry paths in the irreducible BZ1. (c) Close view of the avoided crossing region where the unconventional Rashba spin splitting is observed.

### Band structure with orbital character

The results of the total band structure can be projected onto the contribution of each orbital, referred as orbital projection. The user can obtain it in the current calculation by typing the following:

$ *python BP-bands.py -id run2 -ph MGX -kp 1000 -tk 1 -nb 50 -xy -0.25 0.5 -0.35 0.03*

Now we have selected the M-Γ-X path and *tk* equal to 1 means that the band structure is projected onto the different orbitals in the $t_{2g}$ manifold. The relative weight of the orbitals ($d_{xy}$, $d_{yz}$, $d_{zx}$) is represented by a RGB color code (the user can change the color trio in the *bands.yaml* configuration file). We chose to use a finer discretization along the path with *kp* equal to 1000. The resulting output can be seen in Fig. 6(a).

### Band structure projected onto planes

We can calculate the contribution of a set of planes to the total band structure, referred as plane projection. The user can compute such a projection by typing:

$ *python BP-bands.py -id run2 -ph MGX -kp 1000 -tk 2 -nb 50 -xy -0.25 0.5 -0.35 0.03 -pi 0 -pf 2*

This line indicates the plane projection task by setting *tk* equal to 2. The planes for projections are established by means of the initial plane (*pi*) and final plane (*pf*) parameters, and the set will range from *pi* to *pf*-1. In consequence, the above example will project the total band structure onto the first two planes. The output is shown in Fig. 6(b).

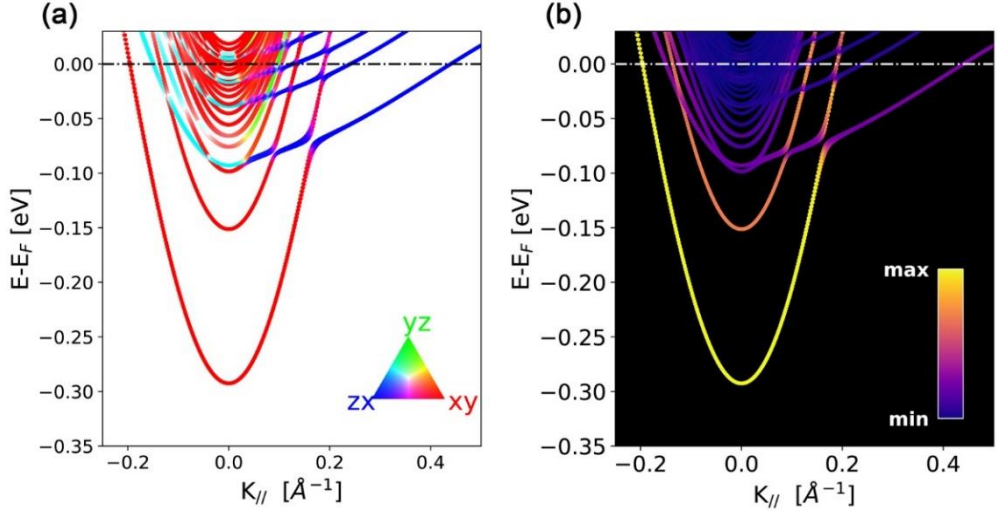

**Fig 6.** Band structure for STO(100) (a) Orbitally projected and (b) projected onto the first two planes along the high symmetry M-Γ-X path.

**4.3.2 Energy slices**

As another part of post-processing, we can compute the set of k-points whose eigenvalues are at a specific energy cut, which is named energy slice. This component can be run by typing:

*$ python BP-energy_slices.py -id run2 -ec 0.0 -nk 300 -ba 60 -bf 0.7*

The energy cut (*ec*) in which the slice will be taken is set to 0.0. Then, we construct a denser k-grid (*nk*) of 300 x 300 points, which will be split in 60 batches (*ba*). The reason for defining batches is the reduction of the Hamiltonian size, in this case to (300 x 300)/60 Hamiltonian matrices, with the subsequent reduction of memory consumption, resulting in the need for less resources and computation time. The k-box factor parameter (*bf*) is to enlarge or reduce the BZ1. By definition, the *ec* equal to 0 is the Fermi level, thus we are computing the Fermi surface. For comparing the results, we can also calculate an energy slice for a different energy cut, for instance, 50 meV below the Fermi level:

*$ python BP-energy_slices.py -id run2 -ec -0.05 -nk 300 -ba 60 -bf 0.7*

Due to the significant computational cost of using a coarse grid, once this component finishes an output file is automatically saved to the identifier folder, then we must apply the *BP-energy_plot.py* component to plot the results. The plots for these two examples are shown in Fig. 7(a) and (b).

It is possible to compute the energy slices for specific regions of the BZ1. In this way, the resolution can be increased at the same computational cost. For example, the unconventional Rashba spin splitting can be now closely examined by typing:

*$ python BP-energy_slices.py -id run2 -ec 0.0 -nk 300 -ba 60 -bf 0.06 -dk 0.09 -0.17*

Note that the k-grid extension is reduced by setting *bf* equal to 0.06. In addition, we are using the k-grid offset parameter (*dk*), which is chosen to be near the area of interest. The output plot associated to this energy slice calculation is shown in Fig 7(c).

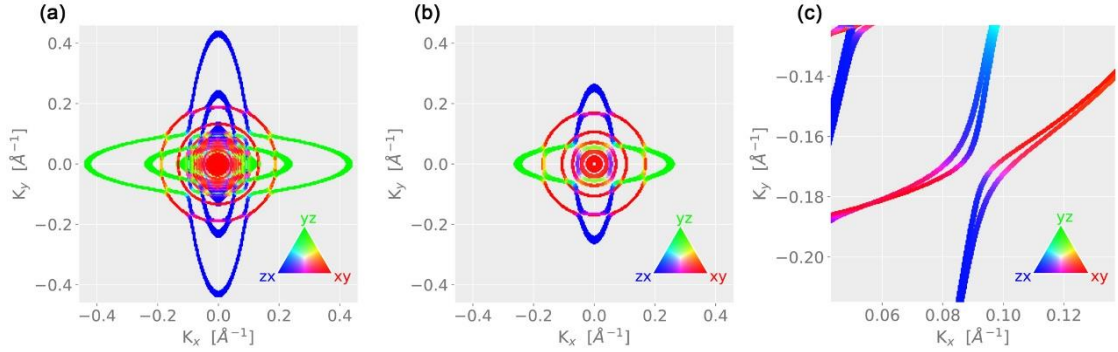

**Fig 7.** Energy slices plot for STO(100) at (a) *ec* = 0.0 (Fermi surface) and (b) *ec* = -0.05 eV. (c) Close up to the avoided crossing region where unconventional Rashba spin splitting is observed.

### 4.3.3 Other capabilities

For the sake of completeness, we show here how to compute the orbital decomposition of the electron density and the envelope wavefunctions of the system. The first quantity allows for analyzing the contribution of the different orbitals to the total electron density. It can be computed by using the *BP-orb_density.py* component as:

*$ python BP-orb_density.py -id run2 -aa 0.3*

where, we set the curves opacity (*aa*) to 0.3. The output is shown in Fig. 8(a), where the total and partial electron densities are indicated accordingly.

Now we want to compute the envelope wavefunctions at the Γ-point. These $r_\perp$-dependent wavefunctions reflect the quantum confinement near the surface or interface. It can be computed by using the *BP-envelope_wfs.py* component and typing:

*$ python BP-envelope_wfs.py -id run2 -sf 0.3 -nw 8 -xy -2 22.5 -0.35 0.06*

here we specify the scale factor (*sf*) which affects the amplitude of the envelope wavefunctions, the number of these wavefunctions (*nw*) to compute and the x and y limits of the plot by means of the *xy* parameter. The output is shown in Fig. 8(b), where the first eight envelope wavefunctions are shown.

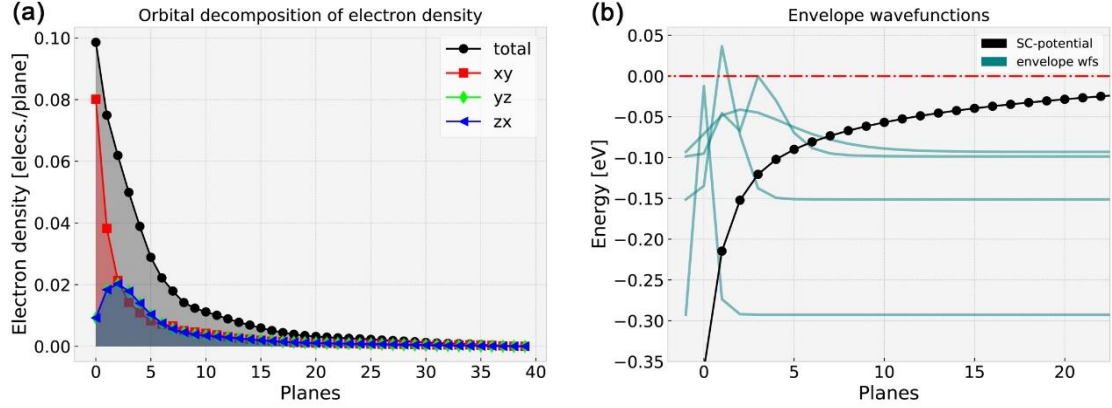

**Fig 8.** Further post-processing steps in STO(100): (a) orbital decomposition of electron density and (b) the first eight envelope wavefunctions.

### 4.4 Comparison with experiments

Several groups have studied experimentally the electronic band structure of the 2DES created at the bare surface of different oxides or at the LAO/STO interface. Here we focus on ARPES [62,63] experiments, since it allows for a direct comparison of the results obtained in *BinPo*. The band structure of the 2DES will have a strong influence in the transport properties of the system and thus, comparisons with experimental results of quantum oscillations [64] and spin transport [31] are possible but require additional modelling and assumptions. In Fig. 9(a) and 9(c) we show measurements of the electronic band dispersion and the ARPES Fermi surface for the 2DES stabilized on the (001) bare surface of STO [8,25,65]. A full description of the measurement conditions can be found in [8]. The energy-momentum dispersions on Fig. 9(a) consist of a ladder of at least three light bands, usually called sub-bands, which are a clear signature of quantum confinement near the surface. Additionally, we observe a single heavy band of only ~50 meV bandwidth in contrast with the ~300 meV in the case of the first light sub-band indicating a strong breaking of the $t_{2g}$ orbital degeneracy present in bulk STO. The observed light and heavy sub-bands contribute with circular and elliptical Fermi surface sheets respectively as observed in Fig. 9(a), in the present case the heavy state appears with very low intensity as a consequence of the geometry of the experiment.

In Fig. 9(b) and 9(d) we show calculations performed with *BinPo* of the Fermi surface and band structure along the X-Γ-X high symmetry direction, respectively. In the calculation the potential energy at the surface top-most layer is a free parameter and determines the bandwidth and total charge in the system. In the present case, it was chosen so that the total bandwidth of the first light sub-band coincides with the experimental value of ~300 meV. There is a good overall agreement with the experimental data for both the electronic band dispersion and the Fermi surface. The relative sub-band energy and Fermi wave vector for these three light bands is well described by the model. As described in Section 4.3.1 above, a plane projection of the band structure indicates that the first sub-bands are spatially confined closer to the surface. Since ARPES is a surface sensitive technique these bands appear with more intensity in the measurements. A similar argument explains why the heavy bands appear less intense in the experiments. However, taking advantage of polarized light it was shown that multiple heavy sub-bands are present in the 2DES as shown in the *BinPo* calculation [66]. We notice that while the effective mass of the light sub-bands is well described by the model there is a clear discrepancy with that of the heavy bands, this mass renormalization due to electron-phonon interaction is not considered in *BinPo* [25].

Using polarized synchrotron light, it can be experimentally demonstrated that the light and heavy bands have predominant $d_{xy}$ and $d_{xz/yz}$ orbital character respectively [9,25,66]. This observation is in agreement with the calculated orbital character shown in Fig. 9(b) where the light bands appear mostly red corresponding to $d_{xy}$ orbitals and the heavy bands appear blue or green depending on whether the orbital character is $d_{xz}$ or $d_{yz}$. Interestingly, at the crossing points of the heavy and light bands the orbital character is mixed and the unconventional Rashba spin splitting is maximized [25,31]. While *BinPo* gives a clear picture of the spin splitting in these so-called avoided crossings, there are experimental reports contradicting these details of the electronic structure [27,67].

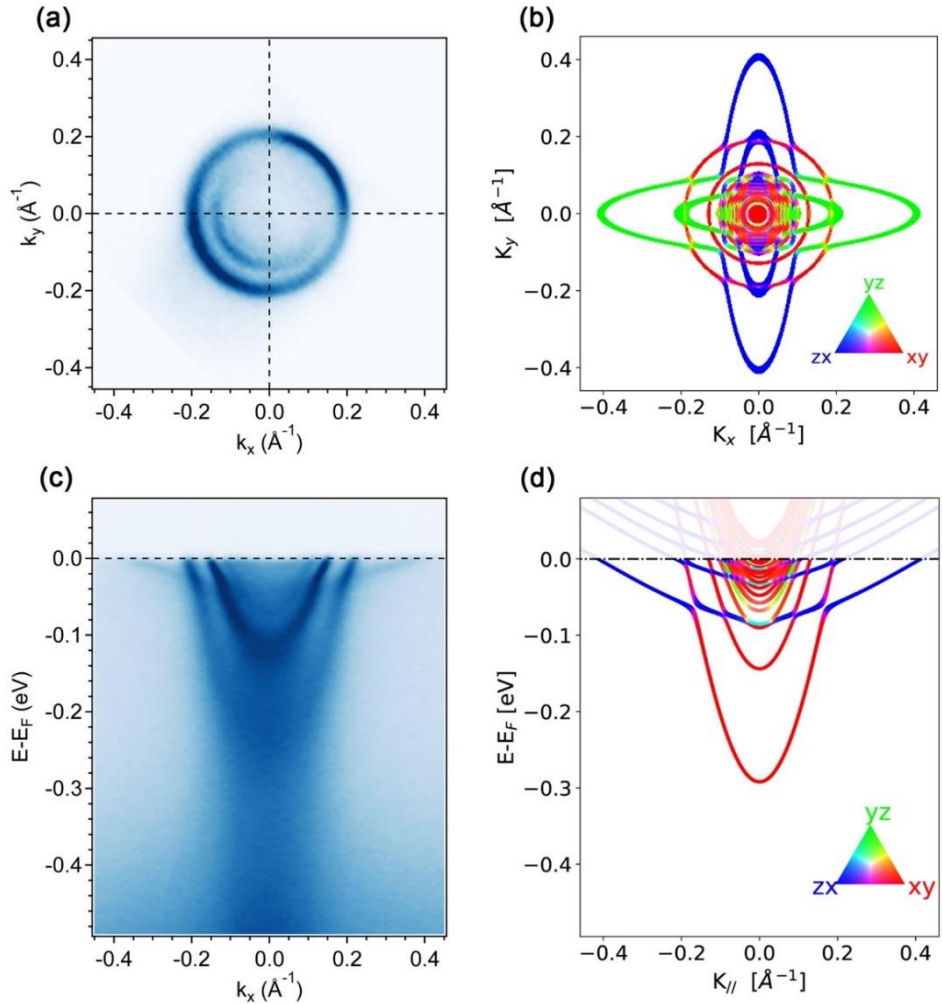

**Fig 9.** ARPES measurement of the 2DES stabilized at the bare STO(100) surface. (a) Fermi surface. (b) *BinPo* simulation of the Fermi surface. (c) Bands dispersions and (d) *BinPo* simulation along X-Γ-X path.

## 5. Summary and conclusions

We have introduced *BinPo*, an open-source code to compute the band structure and other electronic properties of the 2DES generated at the surface or interface of $ABO_3$ perovskites. The code is written in Python 3.x, so it can run in all platforms in a computationally efficient manner. The files required by *BinPo* are based on first-principles calculations (which can be computed by many available codes) followed by the MLWF obtention with the open-source Wannier90 program. We have provided two Wannier files for working with either STO or KTO, but the user can append different files for other cubic $ABO_3$ perovskites or different DFT treatments. We have shown how *BinPo* solves the Schrödinger-Poisson scheme to get the SC potential energy under a reasonable number of assumptions. It can calculate quantities like the band structure, energy slices and envelope wavefunctions among others. The validity of the results, which were all obtained in a conventional laptop computer, is illustrated by direct comparison with ARPES experiments. The ease-of-use of *BinPo* sets the stage for an extended adoption, mostly by the experimental scientists looking for a fast and reliable description of the 2DES band structure, helping to understand the experimental results in these novel materials.

**Declaration of interests**

The authors declare that they have no known competing financial interests or personal relationships that could have appeared to influence the work reported in this paper.

**Acknowledgments**

This work has been supported by Comunidad de Madrid, Spain (Atracción de Talento grant No. 2018-T1/IND-10521) and by MICINN, Spain PID2019-105238GA-I00. We acknowledge Alejandro B. Kolton for fruitful discussions and Felix Baumberger for sharing raw ARPES data used to produce figures 9(a) and 9(c). JIB acknowledges funding through the Santander-UCM project PR87/19.

**Appendix A. Details of DFT calculation and Wannierization**

The DFT calculations were performed using the open-source Quantum Espresso program. We used the projector augmented wave (PAW) method [68] and the standard PBE exchange-correlation (XC) functional [69]. The PAW PBE full-relativistic pseudopotentials were taken from PS Library [70]. For the Ti (Ta) atoms the *3s*, *3p*, *4s*, *3d* (*4d*, *6s*, *5d*) electrons were considered as valence. The Brillouin zone was sampled with a 15 x 15 x 15 and 12 x 12 x 12 k-mesh Monkhorst-Pack grid method for STO and KTO respectively [71]. The cutoff energy for the plane-waves was about 952 eV, whereas for the electron density it was 8160 eV. The convergence threshold for self-consistent field (SCF) calculations was set to 1.36 x $10^{-9}$ eV. The electron smearing was selected as fixed because of the insulating nature of the materials. After performing SCF unit cells calculations and fitting to Murnaghan equation of state [72] we obtained the lattice parameter values $a_{STO}$ = 3.9425 Å and $a_{KTO}$ = 4.0184 Å.

The Wannierization and the calculation of the real space Hamiltonian were done using the open-source Wannier90 software. The KS states were projected onto the $t_{2g}$ manifold, which encompasses the Ti *3d* and the Ta *5d* orbitals. Due to fact that the $t_{2g}$ manifold is an isolated group of bands, a disentanglement is not needed [73]. We use 200 iterations for the minimization procedure to ensure convergence. The difference between consecutive total spreads of Wannier functions was less than $10^{-11}$ Å$^2$.

**Appendix B. Comments on relative permittivity models**

It is worth noting that an important element in Eq. (14) is the field dependent relative permittivity $\varepsilon_r[E]$. Indeed, this introduces the non-linearity of the Poisson equation, and such expressions are of the uttermost importance for solving it. More details on the derivation of the equation can be found in Appendix E. The relative permittivity strongly affects the potential profile along the slab. Copie *et al* [74] have derived an expression for STO(100) at low temperatures, which is used in *BinPo* as a pre-defined model. For the KTO(100) case, we have derived a model at low temperature by fitting the experimental data of Ang *et al* [47] with a sigmoid-like function, as shown in Fig. B.1. The last pre-defined model is the constant model, which is useful to analyze the effect on confinement if there is not electric field dependence in permittivity. Additionally, the user can introduce any relative permittivity model as an input for the calculation.

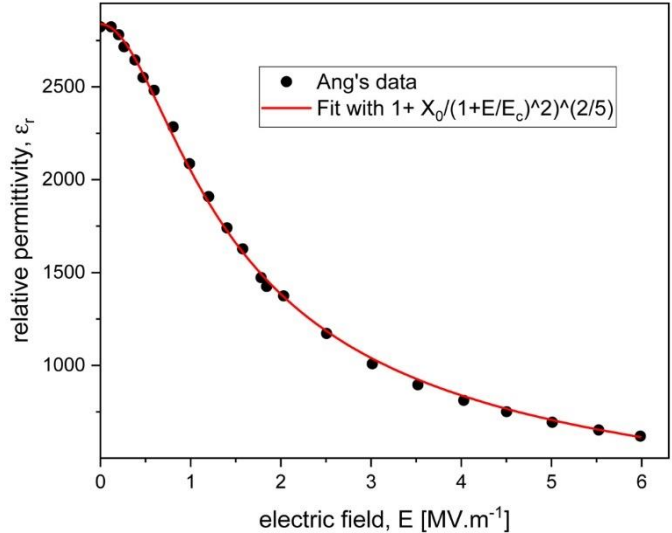

**Fig B.1.** Fit of the experimental Ang's data using a sigmoid-like function. The values obtained from the fitting are $\chi_0 = 2837$ and $E_c = 892244\ V\ m^{-1}$.

We take the functional form of the relative permittivity derived from experimental data as a good starting point. However, an accurate description of relative permittivity as a function of the electric field, and even as a function of the defect density, is a rather intricate problem which is out of the scope of this work. This is the main reason whereby *BinPo* has the facility of using any functional form as input.

**Appendix C. Post-processing calculations**

The eigenstates of Eq. (10) contain useful information of the system through the coefficients $u_{\alpha r_\perp}^{\mathbf{k}_\parallel n}$. So that, once the SC potential is found, other properties can be computed. In particular, we will define two projection operators to get information about the orbital character in the band structure or energy slices, as well as the contribution of different planes to the band structure. In the following, the orthogonality properties of the MLWF basis of Eq. (4) are used to compute the expectation values.

**Orbital projection**

Recalling the orbital index $\alpha$ from Eq. (6), which spans the $t_{2g}$ manifold, we can define the orbital projector onto the orbital $\alpha$ as:

$$\hat{P}_\alpha = \sum_{r_\perp} |\mathbf{R}_\parallel \alpha r_\perp\rangle\langle\mathbf{R}_\parallel \alpha r_\perp| \qquad (C.1)$$

The expectation value of this operator for a given eigenstate, namely, the $\alpha$-orbital character is:

$$C_\alpha^{\mathbf{k}_\parallel n} \equiv \langle\phi_{\mathbf{k}_\parallel n}|\hat{P}_\alpha|\phi_{\mathbf{k}_\parallel n}\rangle = \sum_{r_\perp} \left(u_{\alpha r_\perp}^{\mathbf{k}_\parallel n}\right)^* u_{\alpha r_\perp}^{\mathbf{k}_\parallel n} \qquad (C.2)$$

**Projections onto planes**

We will now define the single plane projector, analogously to Eq. (C.1), as:

$$\hat{\Pi}_{r_\perp} = \sum_\alpha |\mathbf{R}_\parallel \alpha r_\perp\rangle\langle\mathbf{R}_\parallel \alpha r_\perp| \qquad (C.3)$$

Note that in this case the expectation value of $\hat{\Pi}_{r_\perp}$ points out the contribution of the plane at $r_\perp$ to the total band structure. We may be interested in analyzing the contribution of a consecutive set of planes rather than the contribution of a single plane. So, we define the plane projector of a set as $\hat{P}_{r_\perp',r_\perp''} = \sum_{r_\perp} \hat{\Pi}_{r_\perp}$ with $r_\perp$ in the interval $[r_\perp', r_\perp'' - 1] \subseteq [0, L-1]$. The expectation value for a given eigenstate is:

$$C_{r_\perp',r_\perp''}^{\mathbf{k}_\parallel n} \equiv \langle\phi_{\mathbf{k}_\parallel n}|\hat{P}_{r_\perp',r_\perp''}|\phi_{\mathbf{k}_\parallel n}\rangle = \sum_{r_\perp = r_\perp'}^{r_\perp''} \sum_\alpha \left(u_{\alpha r_\perp}^{\mathbf{k}_\parallel n}\right)^* \left(u_{\alpha r_\perp}^{\mathbf{k}_\parallel n}\right) \qquad (C.4)$$

**Orbital decomposition of electron density**

It is useful to find out the partial contribution of the orbitals to the charge density ($\rho(r_\perp)$) along the slab. By retaining the index $\alpha$ from Eq. (12) it is possible to get such a quantity $\rho_\alpha(r_\perp)$ defined by:

$$\rho_\alpha(r_\perp) = -\frac{|e|}{\Omega_\perp N_\parallel} \sum_{\mathbf{k}_\parallel n} f(\epsilon_{\mathbf{k}_\parallel n}, T) \left|u_{\alpha r_\perp}^{\mathbf{k}_\parallel n}\right|^2 \qquad (C.5)$$

**Envelope wavefunctions at the Γ-point**

It is useful to visualize the potential well diagram associated to the SC potential confining the charge. We will be particularly interested at the Γ-point, where the bandwidth is maximum. There, the so-called envelope wavefunctions are defined as:

$$\xi_{\Gamma n}^\alpha(r_\perp) = \sum_\alpha \left(u_{\alpha r_\perp}^{\Gamma n}\right)^* u_{\alpha r_\perp}^{\Gamma n} \qquad (C.6)$$

Note that, despite their different dependences, this equation is similar to Eq. (C.2). In fact, Eq. (C.6) can be easily inferred using a proper projection operator like Eq. (C.1) and (C.3).

## Appendix D. Fixed background density approach

The presence of in-gap states in STO was reported in several works [75–77]. These states are generally associated to defects. When the charge is transferred to the interface and the 2DES emerges, some electrons could be trapped in these localized states. For this reason, we include a minimal approach in *BinPo* to consider such states, whose trapped charge could influence the band structure as well. We follow the approach implemented in other works [78–80]. We will rewrite Eq. (14) with a slight modification:

$$\frac{d^2 V(r_\perp)}{dr_\perp^2} = \frac{|e|}{\varepsilon_0 \varepsilon_r[E(r_\perp)]}\left(\rho(r_\perp) + \rho_{def}(r_\perp)\right) \qquad (D.1)$$

where $\rho_{def}(r_\perp)$ is the profile of fixed charge along the slab, which is an unknown quantity. Our simple approach consists of treating this charge distribution as $\rho_{def}(r_\perp) = \rho_d \Theta(L_d - r_\perp)$, with $\rho_d$ the constant charge per plane, $L_d$ the extension of this charge (from the top-most layers toward the bulk) and Θ the Heaviside step function.

The chance to set the background electron occupation could be accessed by the configuration file for SC potential energy calculation, *~/config_files/scp.yaml*. In Fig. D.1 we show how the band structure is modified when a constant charge of 0.01 electrons per plane along the whole slab is considered. For the calculation shown in Fig. D.1(b), the integrated free charge and fixed charge are approximately the same. The user can reproduce both calculations of Fig. D.1 by looking at *~/BPexamples*.

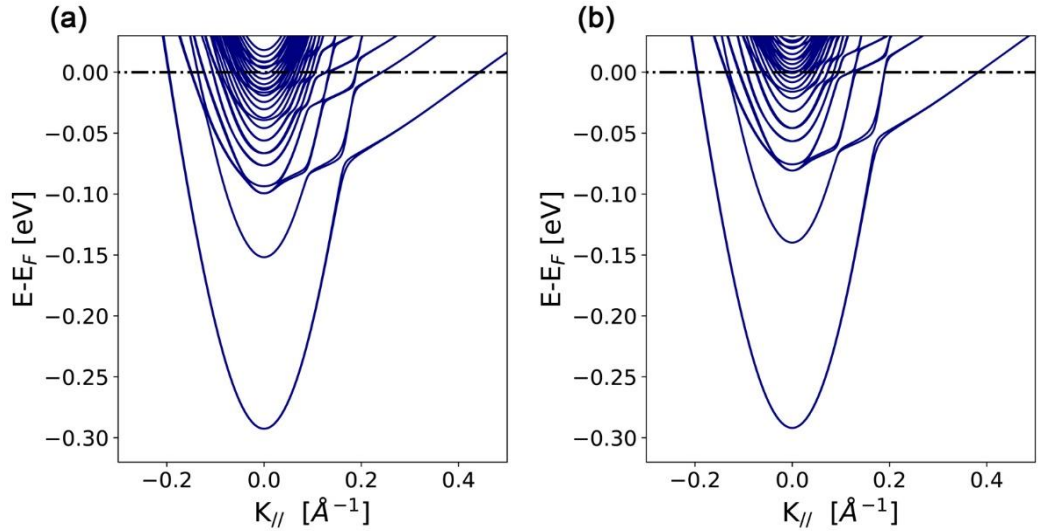

**Fig D.1.** Modification on the band structure inducing by a fixed charge background to simulate defects. (a) STO(100) total band structure along M-Γ-X path. (b) Same as (a) including an additional charge of 0.01 electrons per plane.

## Appendix E. Supplementary Information

# Supplementary Information for *BinPo*: An open-source Python code to compute the band structure of two-dimensional electron systems

Martínez E. A.[a], Beltrán Fínez J. I. [a], Bruno F. Y.[a]

[a] GFMC, Departamento de Física de Materiales, Universidad Complutense de Madrid, 28040 Madrid, Spain

**SI.1 Deduction and validity of the charge density expression**

In order to obtain the expression for the charge density, it would be instructive to start from the bulk case. Consider the bulk Hamiltonian matrix $H^{(\mathbf{k})}$ whose elements are given by Eq. (2) of the main text. Then, the bulk eigenenergies $\varepsilon_{\mathbf{k}n}$, with the associate eigenstates $|\phi_{\mathbf{k}n}\rangle$, can be straightforwardly obtained by diagonalization. In turn, these eigenstates, considering the Eq. (1) of the main text, can be expanded in term of $|\mathcal{W}_{\mathbf{k}\alpha}\rangle$ states [1] as:

$$|\phi_{\mathbf{k}n}\rangle = \sum_\alpha u_\alpha^{\mathbf{k}n} |\mathcal{W}_{\mathbf{k}\alpha}\rangle \qquad (SI.1)$$

where $\alpha$ spans the $t_{2g}$ manifold $\{d_{yz}^\uparrow, d_{yz}^\downarrow, d_{zx}^\uparrow, d_{zx}^\downarrow, d_{xy}^\uparrow, d_{xy}^\downarrow\}$ and the resulting normalization condition reads $\sum_\alpha |u_\alpha^{\mathbf{k}n}|^2 = 1$. We can compute the 3D electron density, $\rho_b(\mathbf{r})$, directly from the eigenstates as:

$$\rho_b(\mathbf{r}) = -|e| \sum_{\mathbf{k}n} f(\varepsilon_{\mathbf{k}n}, T) |\langle\mathbf{r}|\phi_{\mathbf{k}n}\rangle|^2 \qquad (SI.2)$$

Now, we calculate the mean electron density in the bulk by averaging over all space. By the orthogonalization properties of the MLWF basis, we find:

$$\bar{\rho}_b = \frac{1}{N\Omega}\int_\Omega d\mathbf{r}\,\rho_b(\mathbf{r}) = -\frac{|e|}{N\Omega}\sum_{\mathbf{k}n} f(\varepsilon_{\mathbf{k}n}, T) \sum_\alpha |u_\alpha^{\mathbf{k}n}|^2 = -\frac{|e|}{N\Omega} \sum_{occup.} 1 \qquad (SI.3)$$

where $\Omega = a^3$ is the volume of the cubic cell, with $a$ being the lattice parameter and $N$ is the number of mesh points in the first Brillouin zone. It has to be noticed that the last term in Eq. (SI.3) contains the total number of occupied states, so that it returns an intuitive result.

Following the same arguments, we could try to compute the electron density for the slab system and the corresponding average bulk density. In analogy with Eq. (SI.2) we could write:

$$\rho_s(\mathbf{r}) = -|e| \sum_{\mathbf{k}_\| n} f(\varepsilon_{\mathbf{k}_\| n}, T) |\langle\mathbf{r}|\phi_{\mathbf{k}_\| n}\rangle|^2 \qquad (SI.4)$$

$$\bar{\rho}_s = \frac{1}{N_\|\Omega_\perp}\int_{\Omega_\perp} d\mathbf{r}\,\rho_s(\mathbf{r}) = -\frac{|e|}{N_\|\Omega_\perp}\sum_{\mathbf{k}_\| n} f(\varepsilon_{\mathbf{k}_\| n}, T) \sum_{\alpha, r_\perp} |u_{\alpha r_\perp}^{\mathbf{k}_\| n}|^2 \qquad (SI.5)$$

where $\Omega_\perp$ is the interplanar volume as defined in Eq. (13) of the main text and $N_\|$ the mesh points in the 2D first Brillouin zone. Note that the last sum could be replaced by 1 according to the eigenstates normalization. Nevertheless, we want to define a plane dependent electron density. By retaining the $r_\perp$ index in the last term, the electron density can be defined as:

$$\rho(r_\perp) \equiv \bar{\rho}_s^{(r_\perp)} = -\frac{|e|}{N_\| \Omega_\perp} \sum_{\mathbf{k}_\| n} f(\varepsilon_{\mathbf{k}_\| n}, T) \sum_\alpha \left| u_{\alpha r_\perp}^{\mathbf{k}_\| n} \right|^2 \tag{SI.6}$$

that is finally the Eq. (12) of the main text.

It is important to notice that along the {100} orientations, the interplanar and the bulk unit cell volumes are the same, *i. e.*, $\Omega_\perp = \Omega$. Thus, Eq. (SI.6) is exact because of orthogonalization relations in Eq. (4) of the main text are satisfied. Although along other orientations the orthogonalization relations are certainly not true, we could make the following assumption:

$$\langle \mathbf{R}'_\| \alpha' r'_\perp | \mathbf{R}_\| \alpha r_\perp \rangle \cong \delta_{\mathbf{R}_\| \mathbf{R}'_\|} \delta_{\alpha \alpha'} \delta_{r_\perp r'_\perp}$$
$$\langle \mathcal{W}_{\mathbf{k}_\| \alpha' r'_\perp} | \mathcal{W}_{\mathbf{k}_\| \alpha r_\perp} \rangle \cong \delta_{\mathbf{k}_\| \mathbf{k}'_\|} \delta_{\alpha \alpha'} \delta_{r_\perp r'_\perp} \tag{SI.7}$$

We do not compute the overlap between different elements of the maximally localized Wannier functions (MLWF) basis using the interplanar volume. The justification for such relations is based on the strong localization of the elements of the MLWF basis, which is a necessary condition in any tight binding (TB) approach. So that, assuming valid Eq. (SI.7), the electron density as function of planes can be computed using Eq. (SI.6) for any arbitrary crystallographic direction.

### SI.2 Non-linear 1D Poisson equation

**Deduction of the equation**

In non-linear media, the relation between the electric field $E$ and the electric displacement $D$ is $dD = \varepsilon_0 \varepsilon_r(E) dE$, where $\varepsilon_r(E)$ is the field dependent relative permittivity, whose nature is differential. An interesting discussion about the physical meaning of the implied quantities can be found elsewhere [2].

In order to obtain the Poisson equation, we should introduce an expression for $D$ in the Gauss law, $\frac{dD}{dr_\perp} = \rho$, but the differential relation $dD = \varepsilon_0 \varepsilon_r(E) dE$ must be linearized. We can define the average relative permittivity as:

$$\bar{\varepsilon}(E) \equiv \frac{1}{E} \int_E \varepsilon_r(E') dE' \tag{SI.8}$$

By definition, $\bar{\varepsilon}(E)$ satisfies the linear relation $D = \varepsilon_0 \bar{\varepsilon}(E) E$, in consequence, we can now replace $D$ in the Gauss law:

$$\frac{dD}{dr_\perp} = \frac{d}{dr_\perp}(\varepsilon_0 \bar{\varepsilon}(E) E) = \frac{d}{dr_\perp}\left( \varepsilon_0 \int_E \varepsilon_r(E') dE' \right) = \varepsilon_0 \varepsilon_r(E) \frac{dE}{dr_\perp} = \rho \tag{SI.9}$$

Using the relation between the electric field and the potential energy, $E(r_\perp) = \frac{1}{|e|}\frac{dV(r_\perp)}{dr_\perp}$, we finally obtain the Poisson equation:

$$\frac{d^2 V(r_\perp)}{dr_\perp^2} = \frac{|e|}{\varepsilon_0 \varepsilon_r[E(r_\perp)]}\rho(r_\perp) \tag{SI.10}$$

The functional form of $\varepsilon_r[E]$ could arise from theoretical considerations or from phenomenological fittings of experimental and theoretical data. For STO(100) the model derived by Copie *et al* [3] is used, where for KTO(100) we have derived a similar functional form by fitting the Ang's experimental data [4]. There are a few works in the literature that could be useful to either use or determine these empirical models (see Refs. [5–13]).

**Comments on boundary conditions**

Given that Eq. (SI.10) is a second order differential equation, we must explicit two boundary conditions. *BinPo* allows for using Dirichlet boundary conditions at top- and bottom-most layers, *i. e.*, to set the values $V(0)$ and $V(L-1)$. It is also available to use the mixing Dirichlet-Neumann boundary conditions, namely, to set the values $V(0)$ and $\frac{d}{dr_\perp}V(L-1) = 0$. The latter one is especially useful in two situations: (a) if it is desired to truly know the value of convergence for $V(L-1)$ when using a number of planes large enough, or (b) if it is observed that under the present conditions the potential is confined enough to quickly overcome some $V(L-1)$ used as Dirichlet boundary condition. This last case will produce a non-physical behavior of the potential energy profile along the slab. Information about the implementation of the boundary conditions can be read in the next section.

**SI.3 Solving the non-linear 1D Poisson equation by finite differences method**

The finite differences method (FDM) [14] is, along with finite elements, one of the most popular numerical methods to solve differential equations. We have implemented an iterative solver for Poisson equation in *BinPo*. The derivatives of the unknown function, $V(r_\perp)$ in this case, are approximated as finite differences. Let $\ell$ label the planes perpendicular to the (*hkl*) direction, the first and second derivatives of $V(r_\perp)$ under this approximation reads:

$$\frac{dV(r_\perp)}{dr_\perp} \approx \frac{1}{\Delta r_\perp}\left(V_{\ell+\frac{1}{2}} - V_{\ell-\frac{1}{2}}\right) \tag{SI.11}$$

$$\frac{d^2 V(r_\perp)}{dr_\perp^2} \approx \frac{1}{\Delta r_\perp^2}(V_{\ell+1} + V_{\ell-1} - 2V_\ell) \tag{SI.12}$$

where $\ell$ spans the discrete integer list $[0, L-1]$, being $L$ the number of planes in the slab. In Eq. (SI.11) the values $V_{\ell\pm\frac{1}{2}}$ are found by cubic interpolation. Note that due to the discretization applied to this problem, $\Delta r_\perp$ is the interplanar distance for the cubic cell. We will take advantage of the expression of the relative permittivity in terms of electric field, E, which in turn is proportional to the first derivative of $V(r_\perp)$. Replacing the derivatives, the approximate Poisson equation reads:

$$\frac{1}{\Delta r_\perp^2}(V_{\ell+1} + V_{\ell-1} - 2V_\ell) \approx \frac{|e|}{\varepsilon_0 \varepsilon_r[E_\ell]}\rho_\ell \qquad (SI.13)$$

with $E_\ell = \frac{1}{|e|\Delta r_\perp}\left(V_{\ell+\frac{1}{2}} - V_{\ell-\frac{1}{2}}\right)$.

Note that we use the notation $\varepsilon_r[E_\ell]$ to emphasize that the relative permittivity is a functional of the electric field $E_\ell$. Generally, for any initial guess of $V(r_\perp)$, Eq. (SI.13) will not be satisfied. Thus, an equality in this equation must be enforced. If we reorder the terms and introduce the iteration index $j$, it is possible to express the $\ell$-th element of the potential energy for the $(j + 1)$-th iteration as:

$$V_\ell^{j+1} = \frac{1}{2}\left(V_{\ell+1}^j + V_{\ell-1}^j - \frac{\Delta r_\perp^2 |e|\rho_\ell}{\varepsilon_0 \varepsilon_r[E_\ell]}\right) \qquad (SI.14)$$

By many iterations over all $\ell$'s (except for the boundaries, whose values are determined by the boundary conditions), the numerical solution of Poisson equation can be found.

As mentioned above, the iteration procedure in Eq. (SI.14) does not need to consider the potential at the boundaries. In the case of Dirichlet's boundary conditions, $V_0 = bc1$ and $V_{L-1} = bc2$ are set, where *bc1* and *bc2* are input values. For mixing Dirichlet-Neumann boundary conditions instead, $V_0 = bc1$ and $V_{L-1} = V_{L-2}$ are used, where the second condition is obtained straightforwardly by the simple backward difference expression of $\frac{d}{dr_\perp}V(L-1) = 0$ in FDM. The potentials between the $j$-th and $(j + 1)$-th are evaluated according to:

$$\chi_P^2 = \frac{1}{L}\sum_{l=0}^{L-1}\left(\frac{V_\ell^{j+1} - V_\ell^j}{V_0}\right)^2 \qquad (SI.15)$$

Finally, the solution is found where $\chi_P^2$ is less than a convergence threshold. In our case we set the condition $\chi_P^2 < 3 \times 10^{-12}$.

We did not demonstrate theoretically the convergence of this method, but we observed a well convergent behavior for multiple cases with the predefined relative permittivity models, as well as many others that were introduced as input in *BinPo*.

**SI.4 Convergence analysis**

In *BinPo*, for a given material (W90 file) and a confinement direction, the SC solution is determined by the value and type of the boundary conditions, number of planes, convergence threshold for potential, relative permittivity model and, of course, the k-grid sampling. We want to show the convergence analysis for one of the most relevant parameters: the number of points in the k-grid sampling ($N_k \times N_k$). Actually, we will talk about $N_k$, which is the input parameter. We will also assume that the k-grid offset was already settled. Ideally, $N_k \to \infty$ is desired because it will give a quantitative better result that the code can provide within the approximations considered. However, in order to reduce the computational cost, we need to limit the $N_k$ value, increasing the result uncertainty. This is a rather common problem in numerical simulations and a criterion must be taken to decide when the solution is accurate enough. For example, if we take $N_k = 106$ as reference of converged SC potential, we can compute the error between this potential and the other SC potentials (Fig. SI.1(a) displays all the SC potentials) by means of Eq. (15) of the main text. We can look for the $N_k$ that produces an error less than the convergence threshold used in calculations. Following this criterion, we can see that a value $N_k > 26$ we already are under the threshold (see Fig. SI.1(b)), but we take $N_k^{conv} = 36$ for greater assurance. Of course, other criteria can be used. For a rapid calculation, the user could set values for $N_k < N_k^{conv}$. However, if an accurate SC solution is desirable, user should set $N_k \geq N_k^{conv}$. Ultimately, we want to remark that the same analysis could be done for the number of planes and with difference convergence thresholds if a clear criterion is settled. The criteria could be imposed also over the electron density if this quantity is of major interest. The details for all these calculations can be found in *~/BPexamples*.

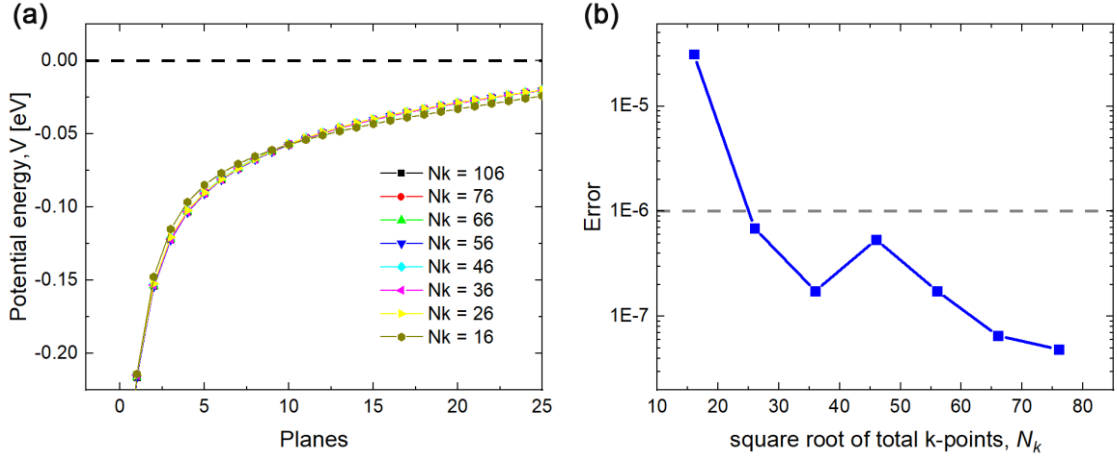

**Fig SI.1.** (a) SC solutions for different values of $N_k$ (plotted up to 25 planes). (b) Error computed among the SC potential and the reference (SC potential with $N_k = 106$).

### SI.5 Change of basis + rotation algorithm for other crystal faces

The 3D bulk TB Hamiltonian actually holds all the information needed to inspect other crystal faces defined by a generic $\vec{v} = [hkl]$ normal direction. We introduce here the simple algorithm to create slabs and compute quantities along arbitrary crystallographic directions. It consists of two steps: (i) change of basis from the original cartesian to the new one, (ii) rotation of the new $\vec{v}$ vector into the [001] direction.

**(i) Change of basis**

It is proper to write explicitly the exponential factor of the 2D Fourier transform in Eq. (3) of the main text, which contains the crystallographic information. First of all, the actual expression used inside the code must be pointed out:

$$e^{i\mathbf{k}_\parallel \cdot \mathbf{R}_\parallel} \rightarrow e^{i 2\pi \mathbf{k}_\parallel \cdot \mathbf{R}_\parallel} \tag{SI.16}$$

with real and reciprocal vectors given by:

$$\mathbf{R}_\| = r_1 \mathbf{a_1} + r_2 \mathbf{a_2}, \quad \mathbf{k}_\| = k_1 \mathbf{b_1} + k_2 \mathbf{b_2} \quad \text{and} \quad \mathbf{a}_i \cdot \mathbf{b}_j = \delta_{ij} \qquad \text{(SI. 17)}$$

with $r_i, k_j$ real coefficients. Note that for $\vec{v} = [001]$, the basis $\mathbf{B} = \{\mathbf{a_1}, \mathbf{a_2}\}$ is the defined by $\{[1,0,0], [0,1,0]\}$. Consider now $\vec{v}$ in an arbitrary direction, thus, a new basis $\mathbf{B'} = \{\mathbf{a_1'}, \mathbf{a_2'}\}$ for the 2D real space and the associated reciprocal basis $\underline{\mathbf{B}_k}' = \{\mathbf{b_1'}, \mathbf{b_2'}\}$ must be considered. Now, we can write arbitrary real and reciprocal vectors analogously to Eq. (SI.17). Let the new coefficients be $\tilde{r}_i, \tilde{k}_j$ then the exponent in Eq. (SI.16) is invariant under changes of basis, as proven below:

$$e^{i2\pi \mathbf{k}_\|'\cdot \mathbf{R}_\|'} = e^{i2\pi(\tilde{k}_1 \mathbf{b_1'} + \tilde{k}_2 \mathbf{b_2'})\cdot(\tilde{r}_1 \mathbf{a_1'} + \tilde{r}_2 \mathbf{a_2'})} = e^{i2\pi(\tilde{k}_1 \cdot \tilde{r}_1 + \tilde{k}_2 \cdot \tilde{r}_2)} = e^{i2\pi \mathbf{k}_\| \cdot \mathbf{R}_\|} \qquad \text{(SI. 18)}$$

where the last equality can be written because the arbitrary indices $\tilde{r}_i, \tilde{k}_j$ span the same discretized real and reciprocal spaces.

**(ii)   Rotation**

So far, we have seen that a change of basis does not affect the results when the Fourier transform is performed. However, it should be noticed that the new transfer integrals must be carefully assigned to certain k-grid. To illustrate this point, we show an example computed with the energy slice post-processing tool for STO(111) as illustrated in Fig. SI.2. At the top we represented schematically the k-grid. Once you have computed the energy slice for the [111] direction, one can observe the distorted energy slice resulting from it (bottom side of Fig. SI.2(a)). This requires rotating the new normal vector into the [001] direction which then produces the correct assignment of k-space symmetries and values (see Fig. SI.2(b)).

As a summary, the Fig. SI.3 shows the simple flowchart of the complete rotation algorithm.

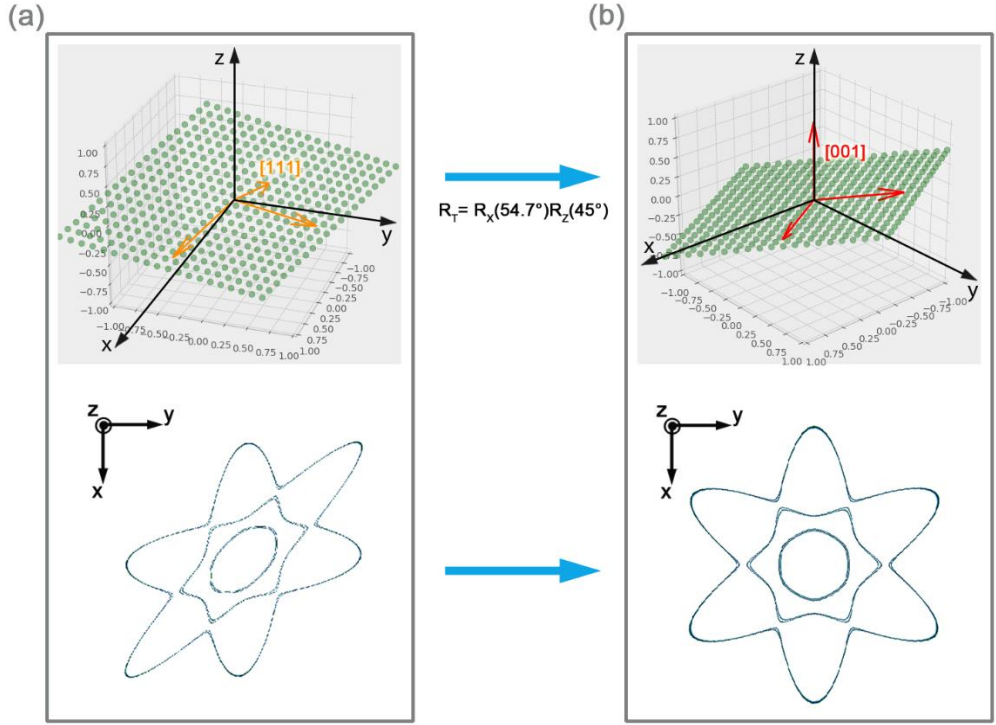

**Fig SI.2.** Illustration of the method to compute the energy slices in other confinement directions: (a) once the change of basis was applied to compute the slice and (b) after applying the 3D rotation that solves the distorsions.

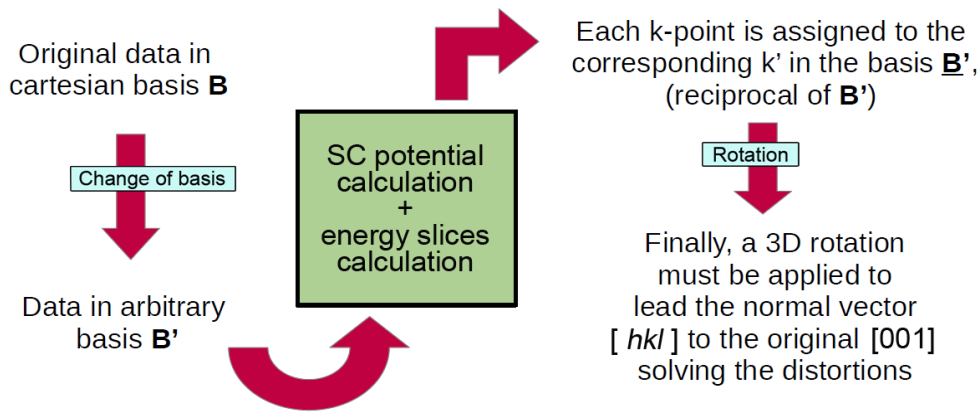

**Fig SI.3.** Flowchart of the algorithm of change of basis + rotation to analyze other faces.